\documentclass[preprintnumbers,amsmath,amssymb,floatfix,12pt,prd,superscriptaddress,nofootinbib]{revtex4}
\usepackage{graphicx}
\usepackage{epsfig}
\usepackage{bm}
\usepackage{amsfonts}

\begin{document}

\title{Thermodynamics of interacting entropy-corrected holographic dark energy
in a non-flat FRW universe}

\author{Mubasher Jamil}
\email{mjamil@camp.nust.edu.pk} \affiliation{Center for Advanced
Mathematics and Physics, National University of Sciences and
Technology, H-12, Islamabad, Pakistan}

\author{Ahmad Sheykhi}
\email{sheykhi@mail.uk.ac.ir} \affiliation{Department of Physics,
Shahid Bahonar University, P.O. Box 76175, Kerman, Iran}

\author{M. Umar Farooq}
\email{mfarooq@camp.nust.edu.pk}\affiliation{Center for Advanced
Mathematics and Physics, National University of Sciences and
Technology,  H-12, Islamabad, Pakistan}

\begin{abstract}
\vspace*{1.5cm} \centerline{\bf Abstract} \vspace*{.5cm} A
so-called ``entropy-corrected holographic dark energy'' (ECHDE),
was recently proposed to explain the dark energy-dominated
universe with the help of quantum corrections to the entropy-area
relation in the setup of loop quantum cosmology. Using this new
definition, we investigate its thermodynamical features including
entropy and energy conservation. We describe the thermodynamical
interpretation of the interaction between ECHDE and dark matter in
a non-flat universe. We obtain a relation between the interaction
term of the dark components and thermal fluctuation. Our study
further generalizes the earlier works [M.R. Setare and E.C.
Vagenas, Phys. Lett. B 666 (2008) 111; B. Wang et al., Phys. Lett.
B 662 (2008) 1] in this direction.
\end{abstract}
\maketitle

\newpage
\section{Introduction}
There is a wide consensus among cosmologists that our universe has
entered a phase of accelerated expansion likely driven by some
unknown energy component whose main feature is to possess a
negative pressure \cite{Rie}. Although the nature of such dark
energy is still speculative, an overwhelming flood of papers has
appeared which attempt to describe it by devising a great variety
of models. Among them are cosmological constant, exotic fields
such as phantom or quintessence, modified gravity, etc, see
\cite{Pad} for a recent review.

An interesting attempt for probing the nature of dark energy within
the framework of quantum gravity, is the so-called ``Holographic
Dark Energy" (HDE) proposal. This model which has arisen a lot of
enthusiasm recently \cite{Coh,Hsu,Li,Huang,HDE,Setare1,wang0}, is
motivated from the holographic hypothesis \cite{Suss1} and has been
tested and constrained by various astronomical observations
\cite{Xin}. The definition of HDE is originally motivated from the
entropy-area relation which depends on the theory of gravity under
consideration. In the thermodynamics of black hole, there is a
maximum entropy in a box of length $L$, commonly termed, the
Bekenstein-Hawking entropy bound, $S\sim M_p^2L^2$, which scales as
the area of the box $A \sim L^2$ rather than the volume $V \sim
L^3$. Here $M^2_p =(8\pi G)^{-1}$ is the reduced Planck mass and
throughout this Paper we use the units $c =\hbar= 1$. In this
context, Cohen et al. \cite{Coh} suggested that in quantum field
theory a short distance cutoff is related to a long distance cutoff
due to the limit set by formation of a black hole, which results in
an upper bound on the zero-point energy density. In line with this
suggestion, Hsu and Li \cite{Hsu,Li} argued that this energy density
could be viewed as the holographic dark energy density satisfying
$\rho_{\Lambda } =3n^2M^2_p/L^2$, where $L$ is the size of a region
which provides an IR cut-off, and the numerical constant $3n^2$ is
introduced for convenience. It is important to note that in the
literature, various scenarios of HDE have been studied via
considering different system's IR cutoff. In the absence of
interaction between dark matter and dark energy in flat universe, Li
\cite{Li} discussed three choices for the length scale $L$ which is
supposed to provide an IR cutoff. The first choice is the Hubble
radius, $L=H^{-1}$ \cite{Hsu}, which leads to a wrong equation of
state, namely that for dust. The second option is the particle
horizon radius. In this case it is impossible to obtain an
accelerated expansion. Only the third choice, the identification of
$L$ with the radius of the future event horizon gives the desired
result, namely a sufficiently negative equation of state to obtain
an accelerated universe. However, as soon as an interaction between
dark energy and dark matter is taken into account, the first choice,
$L=H^{-1}$, in flat universe, can simultaneously drive accelerated
expansion and solve the coincidence problem \cite{pav1}. It was also
demonstrated that in the presence of an interaction, in a non-flat
universe, the natural choice for IR cutoff could be the apparent
horizon radius \cite{shey1}.

As we mentioned the black hole entropy $S$ plays a crucial role in
the derivation of HDE. Indeed, the definition and derivation of
holographic energy density  ($\rho_{\Lambda } =3n^2M^2_p/L^2$)
depends on the entropy-area relationship $S\sim
 A \sim L^2$ of black holes in Einstein's gravity, where $A \sim L^2$ represents
the area of the horizon. However, this definition can be modified
from the inclusion of quantum effects, motivated from the loop
quantum gravity (LQG). The quantum corrections provided to the
entropy-area relationship leads to the curvature correction in the
Einstein-Hilbert action and vice versa \cite{Zhu}. The corrected
entropy takes the form \cite{modak}
\begin{equation}\label{S}
S=\frac{A}{4G}+\gamma \ln {\frac{A}{4G}}+\beta
\end{equation}
where $\gamma$ and $\beta$ are dimensionless constants of order
unity. The exact values of these constants are not yet determined
and still an open issue in loop quantum cosmology. These
corrections arise in the black hole entropy in LQG due to thermal
equilibrium fluctuations and quantum fluctuations \cite{Rovelli}.
Taking  the corrected entropy-area relation (\ref{S}) into
account, the energy density of the HDE will be modified as well.
On this basis, Wei \cite {wei} proposed
 the energy density of the so-called ``entropy-corrected holographic dark energy''
 (ECHDE) in the form
\begin{equation}\label{rhoS}
\rho _{\Lambda }=3n^2M_{p}^{2}L^{-2}+\gamma L^{-4}\ln
(M_{p}^{2}L^{2})+\beta L^{-4}.
\end{equation}
In the special case  $\gamma=\beta=0$, the above equation yields
the well-known holographic energy density. Since the last two
terms in Eq. (\ref{rhoS}) can be comparable to the first term only
when $L$ is very small, the corrections make sense only at the
early stage of the universe. When the universe becomes large,
ECHDE reduces to the ordinary HDE.

In this Paper we intend to study  thermodynamical interpretation of
the interaction between dark matter and ECHDE for a universe with
spacial curvature. As systems's IR cutoff we shall choose the radius
of the event horizon measured on the sphere of the horizon, defined
as $L=ar(t)$. Using the logarithmic correction to the equilibrium
entropy we will derive an expression for the interaction term in
terms of a thermal fluctuation. This Paper is outlined as follows.
In the next section we consider the thermodynamical picture of the
non-interacting ECHDE in a non-flat universe. In section III, we
extend the thermodynamical description in the case where there is an
interaction term between the dark components. We also present an
expression for the interaction term in terms of a thermal
fluctuation in this section. The last section is devoted to
conclusion.
\section{Entropy corrected HDE in a non-flat universe}

We assume the background spacetime to be spatially homogeneous and
isotropic given by Friedmann-Robertson-Walker (FRW) metric
\begin{equation}
ds^2=dt^2-a^2(t)\left[
\frac{dr^2}{1-kr^2}+r^2(d\theta^2+\sin^2\theta d\phi^2) \right],
\end{equation}
where $a(t)$ is the dimensionless scale factor which is an
arbitrary function of time, and $k$  is the curvature parameter
with $k = -1, 0, 1$ corresponding to open, flat, and closed
universes, respectively. The Einstein field equation representing
the dynamics of the FRW spacetime can be written as
\begin{equation}
H^2+\frac{k}{a^2}=\frac{1}{3M_p^2}(\rho_\Lambda+\rho_m),
\end{equation}
where $H=\dot{a}/a$ is the Hubble parameter and $\rho_\Lambda$ and
$\rho_{m}$ are the energy densities of dark energy and dark
matter, respectively. One can rewrite Eq. (4) in the dimensionless
form as
\begin{equation}
1+\Omega_k=\Omega_\Lambda+\Omega_m,
\end{equation}
where the above density parameters are defined by
\begin{equation}
\Omega_m=\frac{\rho_m}{\rho_{cr}}=\frac{\rho_m}{3H^2M_p^2},\ \
\Omega_\Lambda=\frac{\rho_\Lambda}{\rho_{cr}}=\frac{\rho_\Lambda}{3H^2M_p^2},\
\ \Omega_k=\frac{k}{(aH)^2}.
\end{equation}
Here $\rho_{cr}$ is the critical energy density. The energy
conservation equations for dark energy and matter are
\begin{eqnarray}
\dot{\rho}_\Lambda+3H(1+w_\Lambda^0)\rho_\Lambda&=&0,\\
\dot{\rho}_m+3H\rho_m&=&0,
\end{eqnarray}
Notice that in the above equation, we have assumed matter to be
pressure-less fluid, $w_m=0$. As we shall see later, this may not
be the case if matter interacts with dark energy.

Due to deep insights of Bekenstein \cite{beken} and Hawking
\cite{hawking}, there has been established a connection between
gravity and thermodynamics. With the help of semi-classical analysis
of static black holes with a vacuum background, Hawking proposed
that black holes could not remain static but can lose mass by
emitting virtual particles from their horizons. From the pure
relativistic point of view, black holes are `cold' objects without
any temperature. But Hawking's analysis implied that black holes are
not only `hot' but can gradually increase their temperature with
evaporation. Thus a black hole can explode if its mass is reduced
closer to Planck's mass. The analysis showed that the black hole's
horizon temperature is inversely proportional to its mass.
Bekenstein \cite{beken} thought that since the entropy of the black
hole horizon is proportional to horizon area, hence with
evaporation, the black hole's entropy will decrease with time and
violate the second law of thermodynamics. As a black hole loses
entropy, the entropy of the background universe increases.
Bekenstein suggested that the sum of black hole entropy and the
background entropy must be an increasing quantity with respect to
time. It was a matter of doubt whether the temperature associated
with the horizon is a physical temperature or merely a geometrical
effect. To answer this, Padmanabhan \cite{paddy1} proved that this
temperature is a physical quantity like temperature of physical
objects.

In this Paper, following \cite{wei} we assume the energy density of
the HDE is modified due to the correction terms in the entropy
formula. Hence the energy density of the ECHDE takes the form (2)
where $L$ is defined as
\begin{equation}
L=a(t)\frac{\text{sin}(\sqrt{|k|}y)}{\sqrt{|k|}},\ \ \
y=\frac{R_h}{a(t)},
\end{equation}
where $R_h$ is the size of the future event horizon defined as
\begin{equation}
R_h=a(t)\int\limits_t^\infty\frac{dt^\prime}{a(t^\prime)}=a(t)\int\limits_0
^{r_1}\frac{dr}{\sqrt{1-kr^2}}.
\end{equation}
The last integral has the explicit form as
\begin{equation}\int\limits_0^{r_1}\frac{dr}{\sqrt{1-kr^2}}=\frac{1}{\sqrt{|k|}}
\text{sin}^{-1}(\sqrt{|k|}r_1)=
\begin{cases} \text{sin}^{-1}(r_1) , & \, \,k=+1,\\
             r_1, & \, \,  k=0,\\
             \text{sinh}^{-1}(r_1), & \, \,k=-1,\\
\end{cases}\end{equation}
Using the definitions of $\Omega_\Lambda$ and $\rho_{cr}$, it is
straightforward to show that
\begin{equation}
HL=\sqrt{\frac{3n^2M_p^2+\gamma L^{-2}\ln(M_p^2L^2)+\beta
L^{-2}}{3M_p^2\Omega_\Lambda^0}}.
\end{equation}
Differentiating $L$ with respect to time $t$ and using (12) yields
\begin{equation}
\dot L=\sqrt{\frac{3n^2M_p^2+\gamma L^{-2}\ln(M_p^2L^2)+\beta
L^{-2}}{3M_p^2\Omega_\Lambda^0}}-\text{cos}(\sqrt{|k|}y),
\end{equation}
where
\begin{equation}\text{cos}(\sqrt{|k|}y)=
\begin{cases} \cos y  & \, \,k=+1,\\
             1 & \, \,  k=0,\\
             \cosh y & \, \,k=-1.\\
\end{cases}\end{equation}
 Differentiating (2) with respect to $t$ gives
\begin{eqnarray}
\dot{\rho} _{\Lambda }&=&\Big[2\gamma L^{-5}-4\gamma L^{-5}\ln
(M_{p}^{2}L^{2})-4\beta
L^{-5}-6n^2M_{p}^{2}L^{-3}\Big]\nonumber\\&&\times\Big[\sqrt{\frac{3n^2M_p^2+\gamma
L^{-2}\ln(M_p^2L^2)+\beta
L^{-2}}{3M_p^2\Omega_\Lambda^0}}-\text{cos}(\sqrt{|k|}y)\Big].
\end{eqnarray}
Making use of (15) in (7) gives
\begin{eqnarray}
w_{\Lambda }^0 &=&-1-\Big(\frac{ 2\gamma L^{-2} -4\gamma L^{-2}\ln
(M_{p}^{2}L^{2})-4\beta L^{-2}-6n^2M_{p}^{2}}{3(3n^2M_{p}^{2}+\gamma
L^{-2}\ln (M_{p}^{2}L^{2})+\beta
L^{-2})}\Big)\nonumber\\&&\times\Big[1-\sqrt{\frac{3M_p^2\Omega_\Lambda^0}{3n^2M_p^2+\gamma
L^{-2}\ln(M_p^2L^2)+\beta L^{-2}}}\text{cos}(\sqrt{|k|}y)\Big].
\end{eqnarray}
The above expression represents the equation of state of the entropy
corrected holographic dark energy.

After the discovery of black hole thermodynamics, people got
interested in understanding whether one can associate
thermodynamics to the cosmological horizons as well similar to
black hole horizons \cite{davies}. There is a fundamental
difference between cosmological and black hole event horizons: in
the former case, the observer lies inside the horizon while in the
later case; the observer is outside the horizon, hence the
temperature measurement by both the observers may be different.
Moreover, if the thermodynamics of horizons is restricted only to
event horizons, then the later quantity does not always exist for
the cosmological spacetimes. Cai and collaborators \cite{cai} have
proposed that one can associate the Hawking temperature to the
horizon of FRW cosmological spacetime if the universe is enclosed
by an apparent horizon (a trapped surface with vanishing
expansion) and not an event horizon. This temperature is inversely
proportional to the size of the apparent horizon.

The FRW universe may contain several cosmic ingredients including
dark energy, dark matter and radiation. Astrophysical observations
suggest that the energy density of dark energy is the dominant
quantity in the universe. In the present analysis, we shall assume
the FRW universe to contain only dark energy and matter.
Astrophysically, the temperature of dark energy and matter could be
different from that of the apparent horizon: if the temperature of
cosmic fluids is hotter than the apparent horizon then heat will
flow outside the horizon and vice versa. It is possible that the
physical system (consisting of dark energy, matter and the apparent
horizon) ultimately reaches the state of thermal equilibrium in a
finite time. Hence one can take the same temperature for the whole
thermodynamic system. In this Paper, we shall assume the dark energy
to be given by the entropy corrected holographic dark energy.

The first law of thermodynamics is defined as
\begin{equation}
TdS_\Lambda=dE_\Lambda +p_\Lambda dV,
\end{equation}
where volume $V$ is given as
\begin{equation}
V=\frac{4\pi}{3}L^3,
\end{equation}
the energy of the holographic dark energy is defined as
\begin{equation}
E_\Lambda =\rho_\Lambda V=\frac{4}{3}\pi (3n^{2}M_{p}^{2}L+\gamma
L^{-1}\ln (M_{p}^{2}L^{2})+\beta L^{-1}),
\end{equation}
and the temperature of the event horizon is given as
\begin{equation}
T=\frac{1}{2\pi L^0}.
\end{equation}
Substituting the aforementioned expressions for the volume, energy
and temperature in Eq. (17) for the case of the non-interacting
ECHDE model, one obtains
\begin{equation}
dS_\Lambda^{(0)}=\frac{8\pi
^{2}}{3}\Big[3n^{2}M_{p}^{2}(1+3w_\Lambda^0)-(\gamma (L^0)^{-2}\ln
(M_{p}^{2}(L^0)^{2})+\beta (L^0)^{-2})(1-3w_\Lambda^0)+2\gamma
(L^0)^{-2}\Big]L^0dL^0,
\end{equation}
where
\begin{eqnarray}
1+3\omega_\Lambda^0 &=&-2-\frac{ 2\gamma (L^0)^{-2} -4\gamma
(L^0)^{-2}\ln (M_{p}^{2}(L^0)^{2})-4\beta
(L^0)^{-2}-6n^2M_{p}^{2}}{3n^2M_{p}^{2}+\gamma (L^0)^{-2}\ln
(M_{p}^{2}L^{2})+\beta
(L^0)^{-2}}\nonumber\\&&\times\Big[1-\sqrt{\frac{3M_p^2\Omega_\Lambda^0}{3n^2M_p^2+\gamma
(L^0)^{-2}\ln(M_p^2L^2)+\beta
(L^0)^{-2}}}\text{cos}(\sqrt{|k|}y)\Big],
\end{eqnarray}
\begin{eqnarray}
1-3\omega_\Lambda^0 &=&4+\frac{ 2\gamma (L^0)^{-2} -4\gamma
(L^0)^{-2}\ln (M_{p}^{2}(L^0)^{2})-4\beta
(L^0)^{-2}-6n^2M_{p}^{2}}{3n^2M_{p}^{2}+\gamma (L^0)^{-2}\ln
(M_{p}^{2}(L^0)^{2})+\beta
(L^0)^{-2}}\nonumber\\&&\times\Big[1-\sqrt{\frac{3M_p^2\Omega_\Lambda^0}{3n^2M_p^2+\gamma
(L^0)^{-2}\ln(M_p^2(L^0)^2)+\beta
(L^0)^{-2}}}\text{cos}(\sqrt{|k|}y)\Big].
\end{eqnarray}

\section{Thermodynamics of interacting ECHDE}

In this section we consider the interaction between ECHDE
$\rho_\Lambda$  and matter $\rho_m$. Given the unknown nature of
both dark energy and dark matter there is nothing in principle
against their mutual interaction and it seems very special that
these two major components in the universe are entirely
independent. Indeed, this possibility has received a lot of
attention recently \cite{Ame,Zim,wang1,shey0} and in particular,
it has been shown that the appropriate coupling between dark
components can influence the perturbation dynamics and the cosmic
microwave background (CMB) spectrum and account for the observed
CMB low $l$ suppression \cite{wang2}. It was shown that in a model
with interaction the structure formation has a different fate as
compared with the non-interacting case \cite{wang2}. It was also
discussed that with strong coupling between dark energy and dark
matter, the matter density perturbation is stronger during the
universe evolution till today, which shows that the interaction
between dark energy and dark matter enhances the clustering of
dark matter perturbation compared to the noninteracting case in
the past. Therefore, the coupling between dark components could be
a major issue to be confronted in studying the physics of dark
energy. However, so long as the nature of these two components
remain unknown it will not be possible to derive the precise form
of the interaction from first principles. Therefore, one has to
assume a specific coupling from the outset \cite{Ads,Ame2} or
determine it from phenomenological requirements \cite{Zim2}.
Thermodynamical description of the interaction (coupling) between
holographic dark energy and dark matter has been studied in
\cite{wang3}.

In the presence of interaction the corresponding conservation
equations are written as
\begin{eqnarray}
\dot{\rho}_\Lambda+3H(1+w_\Lambda)\rho_\Lambda&=&-Q,\\
\dot{\rho}_m+3H\rho_m&=&Q,
\end{eqnarray}
 The energy
conservation equation (24) gives
\begin{eqnarray}
1+3\omega_\Lambda
&=&-2-\frac{Q}{3M_{p}^{2}H^3\Omega_\Lambda}-\Big(\frac{ 2\gamma
L^{-2} -4\gamma L^{-2}\ln (M_{p}^{2}L^{2})-4\beta
L^{-2}-6n^2M_{p}^{2}}{3n^2M_{p}^{2}+\gamma L^{-2}\ln
(M_{p}^{2}L^{2})+\beta
L^{-2}}\Big)\nonumber\\&&\times\Big[1-\sqrt{\frac{3M_p^2\Omega_\Lambda}{3n^2M_p^2+\gamma
L^{-2}\ln(M_p^2L^2)+\beta
L^{-2}}}\text{cos}(\sqrt{|k|}y)\Big].  \nonumber \\
\end{eqnarray}
The effective equations of state for dark energy and matter are
defined by \cite{ali}
\begin{equation}
\omega_\Lambda^\text{eff}=\omega_\Lambda+\frac{\Gamma}{3H},\ \
\omega_m^\text{eff}=-\frac{1}{r_m}\frac{\Gamma}{3H}.
\end{equation}
Here $r_m=\rho_m/\rho_\Lambda$, and
$\Gamma=Q/\rho_\Lambda=3H(1+r_m)$, is the decay rate of dark energy
into matter. Making use of (27) in (24) and (25), we have
\begin{eqnarray}
\dot{\rho}_\Lambda+3H(1+\omega_\Lambda^\text{eff})\rho_\Lambda&=&0,\\
\dot{\rho}_m+3H(1+\omega_m^\text{eff})\rho_m&=&0.
\end{eqnarray}
We add a logarithmic correction term to the entropy
\begin{equation}
S_\Lambda=S_\Lambda^{(0)}+S_\Lambda^{(1)}.
\end{equation}
where
\begin{equation}
S_\Lambda^{(1)}=-\frac{1}{2}\ln(CT^2),
\end{equation}
is the first order correction term to the entropy involving
temperature $T$ and the heat capacity $C$. The later quantity is
defined as
\begin{equation}
C=T\frac{\partial S_\Lambda^{(0)}}{\partial T} .
\end{equation}
To perform this analysis, we assume $\beta=0$.
Now using (31), we have%
\begin{eqnarray}
C=-\frac{8\pi }{3}\Big[3n^{2}M_{p}^{2}(1+3\omega_\Lambda
^0)-(1-3\omega_\Lambda^0 )\gamma (L^0)^{^{-2}}\ln
(M_{p}^{2}(L^0)^{^{2}})+2\gamma (L^0)^{-2}\Big](L^0)^2.
\end{eqnarray}
For the case of the interacting ECHDE model, one obtains
\begin{equation}
dS_\Lambda=8\pi^2n^2M_p^2(1+3\omega_\Lambda)LdL,
\end{equation}
where
\begin{eqnarray}
1+3\omega_\Lambda &=&\frac{1}{8\pi ^{2}n^{2}M_{p}^{2}L}\frac{dS_\Lambda}{dL},\nonumber\\
&=&\frac{1}{8\pi
^{2}n^{2}M_{p}^{2}L}\Big[\frac{dS_\Lambda^{(0)}}{dL}+\frac{dS_\Lambda^{(1)}}{dL}\Big].
\end{eqnarray}
By comparing (26) and (35), we have
\begin{eqnarray}
\frac{Q}{9M_{p}^{2}\Omega _{\Lambda
}H^{3}}&=&-\frac{2}{3}+\frac{1}{3}\Big(\frac{-2\gamma
L^{-2}+4\gamma L^{^{-2}}\ln
(M_{p}^{2}L^{^{2}})+6n^{2}M_{p}^{2}}{3n^{2}M_{p}^{2}+\gamma
L^{^{-2}}\ln
(M_{p}^{2}L^{^{2}})}\Big)\nonumber\\&&\times\Big[1-\sqrt{\frac{3M_p^2\Omega_\Lambda}{3n^2M_p^2+\gamma
L^{-2}\ln(M_p^2L^2)}}\text{cos}(\sqrt{|k|}y)\Big]\nonumber\\&&
-\frac{1
}{9n^2M_p^2}\Big[3n^{2}M_{p}^{2}(1+3\omega_\Lambda)-(1-3\omega_\Lambda
)\gamma (L^0)^{-2}\ln (M_{p}^{2}(L^0)^2)+2\gamma
(L^0)^{-2}\Big]\frac{L^0}{L}\frac{dL^0}{dL}\nonumber\\&&-\frac{1}{24\pi
^{2}n^{2}M_{p}^{2}L}\frac{dS_\Lambda^{(1)}}{dL}.
\end{eqnarray}
In this way we provide the relation between the interaction term
of the dark components and the thermal fluctuation.
\section{Conclusion}
It is interesting to ask whether thermodynamics in an accelerating
universe can reveal some properties of dark energy. It was first
pointed out in \cite{Jac} that the hyperbolic second order partial
differential Einstein equation has a predisposition to the first
law of thermodynamics. The profound connection between the
thermodynamics and the gravitational field equations has also been
observed in the cosmological situations
\cite{Cai2,Cai3,CaiKim,Wang,Cai4,shey2}. This connection implies
that the thermodynamical properties can help understand the dark
energy, which gives strong motivation to study thermodynamics in
the accelerating universe.

On the other side, in the absence of a symmetry that forbids the
interaction between two dark components of the universe there is
nothing, in principle, against it. Further, the interacting dark
mater-dark energy (the latter in the form of a quintessence scalar
field and the former as fermions whose mass depends on the scalar
field) has been investigated at one quantum loop with the result
that the coupling leaves the dark energy potential stable if the
former is of exponential type but it renders it unstable
otherwise. Thus, microphysics seems to allow enough room for the
coupling; however, this point is not fully settled and should be
further investigated.  The difficulty lies, among other things, in
that the very nature of both dark energy and dark matter remains
unknown whence the detailed form of the coupling cannot be
elucidated at this stage.

In this Paper, we investigated the model of interacting holographic
dark energy with the inclusion of entropy corrections to the
holographic dark energy. These corrections are motivated from the
LQG which is one of the promising theories of quantum gravity. We
provided a thermodynamical description of the ECHDE model in a
universe with spacial curvature. We assumed that in the absence of a
coupling, the two dark components remain in separate thermal
equilibrium and that the presence of a small coupling between them
can be described as stable fluctuations around equilibrium. Finally,
resorting to the logarithmic correction to the equilibrium entropy
we derived an expression for the interaction term in terms of a
thermal fluctuation.


\begin{thebibliography}{99}
\bibitem{Rie} A.G. Riess, et al., Astron. J.  116 (1998)
1009;\\
  S. Perlmutter, et al.,  Astrophys. J.  517 (1999) 565;\\
  S. Perlmutter, et al.,  Astrophys. J.  598 (2003) 102;\\
  P. de Bernardis, et al.,  Nature  404 (2000) 955.

\bibitem{Pad} T. Padmanabhan, Phys. Rep.  380 (2003) 235;\\
P. J. E. Peebles,  B. Ratra,  Rev. Mod. Phys. 75 (2003) 559;\\
E.J. Copeland, M. Sami, S. Tsujikawa, Int. J. Mod. Phys. D 15
(2006) 1753.

\bibitem{Coh}  A. Cohen, D. Kaplan, A. Nelson, Phys. Rev. Lett. 82 (1999)
4971.

\bibitem{Hsu} S. D. H. Hsu, Phys. Lett. B 594 (2004) 13.



\bibitem{Li} M. Li, Phys. Lett. B 603 (2004) 1.
\bibitem{Huang} Q. G. Huang, M. Li, JCAP 0408 (2004) 013.


\bibitem{HDE} E. Elizalde, S. Nojiri, S.D.
Odintsov, P. Wang, Phys. Rev. D 71 (2005) 103504;\\  B. Guberina,
R. Horvat, H. Stefancic, JCAP 0505 (2005) 001;\\ B. Guberina, R.
Horvat, H. Nikolic, Phys. Lett. B 636 (2006) 80;\\ H. Li, Z. K.
Guo, Y. Z. Zhang, Int. J. Mod. Phys. D 15 (2006) 869;
\\ Q. G. Huang, Y. Gong, JCAP 0408 (2004) 006;\\
J. P. B. Almeida, J. G. Pereira, Phys. Lett. B 636 (2006) 75;
\\  Y. Gong, Phys. Rev. D 70 (2004) 064029; \\ B. Wang, E.
Abdalla, R. K. Su, Phys. Lett. B 611 (2005) 21.


\bibitem{Setare1}  M. R. Setare, S. Shafei, JCAP 09 (2006) 011;\\
 M. R.  Setare, Phys. Lett. B 644 (2007) 99;\\ M. R. Setare, JCAP
0701 (2007) 023;\\ M. R. Setare, Phys. Lett. B 654 (2007) 1;\\
M. R. Setare, Phys. Lett. B 642  (2006) 421.




\bibitem{wang0} B. Wang, C. Y. Lin and E. Abdalla, Phys. Lett. B 637
(2005) 357;\\ M. R. Setare, Phys. Lett. B 642 (2006)1.



\bibitem{Suss1}  G. 't Hooft, gr-qc/9310026;\\
 L. Susskind, J. Math. Phys. 36 (1995)
6377.

\bibitem{Xin} X. Zhang, F. Q. Wu,  Phys. Rev. D 72 (2005)
043524;\\ X. Zhang, F. Q.  Wu, Phys. Rev. D 76 (2007) 023502;\\
Q. G. Huang, Y.G. Gong, JCAP 08 (2004) 006;\\ K. Enqvist, S.
Hannestad, M. S. Sloth, JCAP 02 (2005) 004;\\ J. Y. Shen, B. Wang,
E. Abdalla, R.K. Su, Phys. Lett. B 609 (2005) 200.

\bibitem{pav1} D. Pavon, W. Zimdahl, Phys. Lett. B 628 (2005)
206.

\bibitem{shey1} A. Sheykhi, Class. Quantum Grav. 27 (2010) 025007.

\bibitem{Zhu} T. Zhu and J-R. Ren, Eur. Phys. J. C 62 (2009) 413;\\ R-G. Cai
et al, Class.Quant.Grav.26:155018,2009

\bibitem{modak} M. Jamil and M. U. Farooq, JCAP 03 (2010) 001;\\
 R. Banerjee and B. R. Majhi, Phys. Lett. B 662 (2008)
62;\\ R. Banerjee and B. R. Majhi, JHEP 0806 (2008) 095;\\ B. R.
Majhi, Phys. Rev. D 79 (2009) 044005;\\ R. Banerjee and S. K. Modak,
JHEP 0905 (2009) 063;\\ S. K. Modak, Phys. Lett. B 671 (2009) 167.

 \bibitem{Rovelli} C. Rovelli,
Phys. Rev. Lett. 77 (1996) 3288;\\ A. Ashtekar, J. Baez, A.
Corichi, and K. Krasnov, Phys. Rev. Lett. 80 (1998) 904;\\ A.
Ghosh and P. Mitra, Phys. Rev. D 71 (2005) 027502;\\ K.A.
Meissner, Class. Quant. Grav. 21 (2004) 5245;\\ A.J.M. Medved and
E.C. Vagenas, Phys. Rev. D 70 (2004) 124021

\bibitem{wei} H. Wei, Commun. Theor. Phys. 52 (2009) 743.

\bibitem{beken} J.D. Bekenstein, Nuovo Cim. Lett. 4 (1972)
737;\\ J.D. Bekenstein, Phys. Rev. D 7 (1973) 2333;\\ J.D.
Bekenstein, Phys. Rev. D 9 (1974) 3292.


\bibitem{hawking} J. M. Bardeen, S. W. Hawking and B. Carter, Comm. Math. Phys.
31 (1973) 161;\\ S. W. Hawking, Commun. Math. Phys. 43 (1975) 199;\\
G. W. Gibbons and S. W. Hawking, Phys. Rev. D 15 (1977) 2738.


\bibitem{paddy1} T. Padmanabhan, Phys. Rept. 406 (2005) 49;\\
T. Padmanabhan, Gen. Relativ. Grav. 40 (2008) 529;\\ T. Padmanabhan,
arXiv:0910.0839v2 [gr-qc];\\ T. Padmanabhan, arXiv:0911.5004v2
[gr-qc].


\bibitem{davies} P.C.W. Davies, Rept. Prog. Phys. 41 (1978) 1313;\\
  P.C.W. Davies, Class. Quant. Grav. 4 (1987 ) L225;\\
  T.M. Davies et al, Class. Quant. Grav. 20 (2003)
2753;\\ H. M. Sadjadi, Phys. Rev. D 73 (2006) 063525;\\ H. M.
Sadjadi and M. Jamil, arXiv:1002.3588v1 [gr-qc];\\ M. Jamil, E. N.
Saridakis and M. R. Setare, Phys. Rev. D 81 (2010) 023007;\\
M. Jamil, E. N. Saridakis and M. R. Setare, arXiv:1003.0876v1
[hep-th].


\bibitem{cai}  R.G. Cai, L.M. Cao, Y.P. Hu, Class. Quantum. Grav. {\bf26} 155018 (2009); \\
 R. Li, J. R. Ren, D. F. Shi, Phys. Lett. B {\bf670} (2009) 446.

\bibitem{Ame} L. Amendola, Phys. Rev. D 60 (1999)  043501; \\ L. Amendola, Phys. Rev. D 62 (2000) 043511;
 \\ L. Amendola and C. Quercellini, Phys. Rev. D 68
(2003)  023514; \\ L. Amendola and D. Tocchini-Valentini, Phys.
Rev. D 64 (2001)  043509.

\bibitem{Zim} W. Zimdahl, D. Pavon, L.P. Chimento, Phys. Lett. B 521 (2001) 133;\\ W. Zimdahl and D. Pavon, Gen. Rel. Grav. 35
(2003) 413;\\ L. P. Chimento, A. S. Jakubi, D. Pavon and W.
Zimdahl, Phys. Rev. D 67 (2003)  083513.


\bibitem{wang1} B. Wang, Y. Gong and E. Abdalla, Phys. Lett. B 624
(2005) 141.


\bibitem{shey0} A. Sheykhi, Phys  Lett  B 681 (2009) 205.
%Interacting holographic dark energy in Brans-Dicke theory%


\bibitem{wang2} B. Wang, J. Zang, Ch.Y. Lin, E. Abdalla, S. Micheletti, Nucl.
Phys. B 778 (2007) 69.




\bibitem{Ads} S. Das, P.S. Corasaniti, J. Khoury, Phys. Rev. D 73 (2006) 083509.
\bibitem{Ame2} L. Amendola,
S. Tsujikawa, M. Sami, Phys. Lett. B 632 (2006) 155;\\ L.
Amendola, C. Quercellini, Phys. Rev. D 68 (2003) 023514;

\bibitem{Zim2} L.P. Chimento, A.S. Jakubi, D. Pavon,W. Zimdahl, Phys. Rev. D 67
(2003) 083513;\\ S. del Campo, R. Herrera, D. Pavón, Phys. Rev. D
70 (2004) 043540.

\bibitem{wang3} B. Wang,  C.Y Lin, D. Pavon, E. Abdalla, Phys. Lett. B 662 (2008)
1;\\ M.R. Setare,  E.C. Vagenas, Phys. Lett. B 666 (2008) 111;\\
A. Sheykhi, M.R. Setare, arXiv:0912.1408.

\bibitem{ali}  M. Jamil and M. A. Rashid, Eur. Phys. J. C 56 (2008) 429;\\
M. Jamil and M. A. Rashid, Eur. Phys. J. C 58 (2008) 111;
\\ M. Jamil and M. A. Rashid, Eur. Phys. J. C 60 (2009) 141;\\
M. Jamil and F. Rahaman, Eur. Phys. J. C 64 (2009) 97;\\
M. Jamil, Int. J. Theor. Phys. 49 (2010) 144.


\bibitem{Jac} T. Jacobson, Phys. Rev. Lett. {\bf75}, 1260 (1995).


\bibitem{Cai2} M.~Akbar and R.~G.~Cai, Phys. Rev. D {\bf 75} 084003
(2007);\\ M. Akbar and R. G. Cai, Phys. Lett. B 648 (2007) 243;\\
M. Akbar and R. G. Cai, Phys. Lett. B 635 (2006) 7;\\ M. Akbar,
Chin. Phys. Lett. 24 (2007) 1158;\\ M. Akbar, Int. J. Theor. Phys.
48 (2009) 2665.




  \bibitem{Cai3} R.~G.~Cai and L.~M.~Cao, Phys.Rev. D {\bf 75}, 064008
  (2007).

\bibitem{CaiKim} R. G. Cai and S. P. Kim, JHEP {\bf0502}, 050
(2005).

 \bibitem{Wang} B. Wang, E.
Abdalla and R. K. Su, Phys.Lett. B {\bf503},  394 (2001);\\ B.
Wang, E. Abdalla and R. K. Su, Mod. Phys. Lett. A {\bf17},  23
(2002);\\ R.~G.~Cai and Y.~S.~Myung, Phys.\ Rev.\ D {\bf 67},
124021 (2003).
\bibitem{Cai4} R.~G.~Cai and L.~M.~Cao,
  Nucl. Phys. B {\bf785} (2007) 135.

\bibitem{shey2} A. Sheykhi, B. Wang and R. G. Cai, Nucl. Phys. B {\bf
779} (2007)1;\\ A. Sheykhi, B. Wang and R. G. Cai, Phys. Rev. D
{\bf 76} (2007) 023515;\\ A. Sheykhi, B. Wang, Phys. Lett. B 678
(2009) 434;\\ A. Sheykhi, JCAP 05 (2009) 019; \\ A. Sheykhi, B.
Wang, Mod. Phys. Lett. A, in press, arXiv:0811.4477.




%%%%%%%%%%%%%%%%%%%%%%%%%%%%%%%%%%%%%%%%%%%%%%%%%%%%%%%%%%%%%%%%%%%%

\end{thebibliography}
\end{document}